\documentstyle[aps,multicol,epsfig,eqsecnum]{revtex}
\newcommand\be{\begin{eqnarray}}
\newcommand\ee{\end{eqnarray}}
\newcommand\ba{\begin{array}}
\newcommand\ea{\end{array}}
\def\r{\rangle}
\def\l{\langle}

\def\T{{\rm Tr}}
\def\cH{{\cal H}}

\def\cI{{\cal I}}

\def\cU{{\cal U}}
\def\cE{{\cal E}}

\def\cD{{\cal D}}

\def\cG{{\cal G}}

\def\cB{{\cal B}}
\def\openone{{\it I}}

\begin{document}
\title{All (qubit) decoherences: Complete characterization and physical implementation}
\author{M\'ario Ziman$^{1,2}$ and Vladim\'\i r Bu\v zek$^{1,3}$}
\address{
$^{1}$Research Center for Quantum Information, Slovak Academy of Sciences,
D\'ubravsk\'a cesta 9, 845 11 Bratislava, Slovakia \\
$^{2}$ {\em Quniverse}, L{\'\i}\v{s}\v{c}ie \'{u}dolie 116, 841 04 Bratislava, Slovakia\\
$^{3}$ Abteilung Quantenphysik, Universit\"{a}t Ulm, 89069 Ulm, Germany
 }
\maketitle

\begin{abstract}
We investigate decoherence channels that are modelled as
a sequence of  collisions of a quantum system (e.g., a qubit) with
particles (e.g., qubits) of the environment. We show that
collisions induce  decoherence when  a bi-partite interaction
between the system qubit and an environment (reservoir) qubit is
described by the controlled-$U$ unitary transformation (gate).
We characterize decoherence channels and in the case of a qubit
we specify the most general decoherence channel and derive a corresponding
master equation. Finally, we analyze entanglement that is generated during the process
of decoherence between the system and its environment.
\end{abstract}

\begin{multicols}{2}
\section{Introduction}
One of the most distinctive features of quantum systems is their ability to ``exist''
in superpositions of mutually exclusive (orthogonal) states \cite{Peres}. Providing a
quantum system has been prepared in a pure state $|\Psi\rangle$ then we can write
$|\Psi\rangle = \sum_k c_k |\psi_k\rangle$, where $|\psi_k\rangle$ are orthonormal vectors that compose
a basis ($\langle\psi_l|\psi_k\rangle=\delta_{kl}$). All bases are unitarily equivalent and
we can express the same state in different bases.
In fact, we can always select a basis such that $|\Psi\rangle$ is a basis
vector so in its matrix representation the vector $|\Psi\rangle$ is represented by a single
diagonal element. According to quantum postulates for the isolated system any evolution is
governed by unitary transformations and the original information about the state preparation
of the quantum system is preserved. As soon as an interaction with
an environment comes into the play (the quantum system is open)
the situation becomes dramatically different and the state is no longer
described by the single diagonal element in some basis.
Depending on properties of the environment
and the character of the interaction our system evolves non-unitarily and its state is, in general, described by
a statistical mixture. Among various possible dynamics of an open quantum system interacting with its environment
a specific role is played by a process in which the off-diagonal elements of the original
state $\varrho=|\Psi\rangle\langle\Psi|$
in some basis are continuously suppressed in time, i.e.
\be
\varrho \to \varrho_{t\to\infty}={\rm diag}[\varrho].
\label{1.1}
\ee
This is a process of {\em decoherence} during which some of the information about the initial state
of the quantum system
might be irreversibly lost \cite{zeh,zurek,zurek2003}.
The basis in which the decoherence takes place is specified by properties
of the environment and the character of the interaction \cite{zurek2003}. There are at least two aspects
of quantum decoherence that keep it in the center of interests in multiple investigation related to foundations
of quantum mechanics and in quantum information processing. The first aspect is, that decoherence is presently
viewed as a mechanism via which classicality emerges from the realm of quantum (see e.g.
\cite{zeh,zurek,zurek2003,giulini,schlosshauser}). In this context it is of paramount importance to specify the basis
(the so called pointer basis \cite{zurek2003}) in which the decoherence takes place.
In the field of quantum information the decoherence is an evil - it degrades quantum resources (superpositions
of states and quantum entanglement) that are needed for quantum information processing \cite{nielsen}.
The degradation of resources is caused by random interactions (errors) between a quantum system under consideration
(e.g. a qubit or a quantum register) with its environment.
If nothing else then these two facets of quantum decoherence are enough to justify an investigation of
decoherence channels (transformations).

As mentioned above the decoherence is caused by
(unavoidable) interactions between the system and its environment.
Consequently, the whole process of decoherence can be completely described
within the framework of the quantum theory as a unitary process that governs
the joint evolution of the quantum systems and its environment \footnote{
Another possibility would be to include decoherence
into the basic dynamical equation, i.e. to add a non-hamiltonian part
into the Schr\"odinger equation \cite{milburn}. However, the modifications of
the basic quantum dynamical law are out of scope of this paper.} \cite{zeh,zurek,zurek2003}.
There are plentiful theoretical models describing the decoherence
within the framework of the standard quantum theory that have been in accordance with various experiments
\cite{haroche,arndt}. These models either use Hamiltonian
evolution of the composite system-plus-enviroment structure
(the Hamiltonian itself is time-independent).

Alternatively, the description of decoherence
can be based on a simple collision-like models, i.e. a sequence of
interactions between the object
under consideration and particles from environment leads to
decoherence.
These models allow us to
study {\it microscopic} dynamics of open systems, in which the flow of
information from the system to the environment and creation of entanglement can be analyzed. In fact,
collision models are equivalent to more general models of causal
memory channels \cite{dennis}. In this case, the memory is
represented by the system under decoherence, whereas the reservoir (environment) plays the role
of input/output systems.

In the present paper we will focus our attention on collision-like
models of decoherence of qubits.
Our first aim is to completely
classify all possible decoherence channels of a qubit. The second task is to show that
all decoherence maps of qubits can be modelled as sequences of collisions.
The paper is organized as follows: Sections II and III
are devoted to a description of general properties of all decoherence channels. In Sec. IV
we present a generic collision-like model. In the
Sec. V the master equations for collision models are derived
and {\it all} possible master equations describing decoherence of a qubit are presented.
In Sec. VI we analyze how entanglement is created during a sequence of collisions.
Finally, in Sec. VII we summarize our results and formulate some open problems.

\section{Decoherence channels}
The aim of this section is to classify all possible completely positive
trace-preserving maps (quantum channels) that describe quantum decoherence.
Let us denote by $\cD$ the set all maps $\cE$
satisfying the decoherence conditions,
i.e.
\be
\l e_k |\cE[\varrho]|e_k\r &=& \l e_k |\varrho|e_k\r  \ \ \ {\rm for\ all}\ \
k
\label{2.1}
\\
|\l e_k|\cE[\varrho]|e_l\r| &<& |\l e_k|\varrho|e_l\r| \ \ \ {\rm for\ all}\ \
k\ne l\, ,
\label{2.2}
\ee
with $\cB=\{|e_k\r\}$ being the {\it decoherence basis}.
For our purposes it is useful to fix one basis $\cB$ and to analyze
all decoherences (forming the set $\cD_\cB$)
with respect to this basis. The general decoherence maps
are then just unitary rotations of elements from $\cD_\cB$, that
correspond to a change of the decoherence basis. In particular, if
$\cE$ is a decoherence map, then also
$\cE^\prime=\cU_1\cE\cU_2$ is such a map. We used the notation
$\cU_j[\varrho]=U_j\varrho U_j^\dagger$ with $U_j$ unitary operators.
>From the definition it is clear that
decoherence channels are unital (they preserve the total mixture, i.e.
$\cE[\openone]=\openone$)
and are not strictly contractive (they might have more than a single fixed point).

Denoting by $\cD_\cB$ the set
of all decoherence maps with respect to a fixed basis $\cB$ we can write
$\cD=\cup_{\cB} \cD_\cB$. Each decoherence map $\cE\in\cD$ belongs only to
one class $\cD_\cB$.
Elements of $\cD_\cB$ and $\cD_{\cB^\prime}$ are unitarily related, i.e.
\be
\nonumber
\cD_{\cB^\prime}=\{\cE^\prime\,|\,\cE^\prime[\varrho]:=\cE [U\varrho U^\dagger], \cE\in\cD_\cB, \cB^\prime=U\cB\}=\cD_{U\cB}\, .
\ee
This defines a new decoherence class only if $\cB^\prime\ne \cB$.
That is, the unitary operation $U$ does not commute with all projectors
$|e_k\r\l e_k|$, or equivalently the basis $\cB$ is not an eigenbasis
of the transformation $U$. If $[U,|e_k\r\l e_k|]=0$ for all $k$
then from a given $\cE\in\cD_\cB$ we obtain different decoherence maps within
the fixed set $\cD_\cB$.

\subsection{Qubit decoherences}

In what follows we will analyze the case of qubit decoherence channels.
In this case the set $\cD$ has surprisingly simple form.
We will use the so-called left-right notation, in which the evolution map
is represented by a 4x4 matrix \cite{ruskai}. Let us choose
the following operator basis
\be
S_0&=&\openone\; ;
\nonumber\\
S_1&=& |\psi\r\l\psi^\perp|+|\psi^\perp\r\l\psi |\; ;
\nonumber\\
S_2&=& i|\psi\r\l\psi^\perp|-i|\psi^\perp\r\l\psi |\; ;
\nonumber\\
S_3&=& |\psi\r\l\psi|-|\psi^\perp\r\l\psi^\perp |\; ,
\ee
where $\cB=\{|\psi\r,|\psi^\perp\r\}$ is the decoherence basis.
The elements of $S$-basis satisfy the same properties as the Pauli
operators, because $S_j=W \sigma_j W^\dagger$ with $W$ being a unitary operation. In this basis
the operators (states) take the form of four-dimensional
vectors $\varrho=\frac{1}{2}(\openone+\vec{r}\cdot \vec{S})
\leftrightarrow \vec{r}_\varrho=(1,\vec{r})$, where $r_j=\T[\varrho
  S_j]$. The evolution $\cE$ is described by 4x4 matrix with elements given by
the equation $\cE_{kl}=\frac{1}{2}\T(S_k\cE[S_l])$. Because of the
trace-preservation we have $\cE_{00}=1$ and $\cE_{01}=\cE_{02}=\cE_{03}=0$.
Consequently, we obtain the Bloch sphere representation
\cite{nielsen} of the state
space, in which the states are illustrated as points (three-dimensional real
vectors $\vec{r}$) lying inside a sphere with a unit radius. The action
of $\cE$ corresponds to an affine transformation of the Bloch vector
$\vec{r}$, i.e. $\vec{r}\to\vec{r}^\prime=T\vec{r}+\vec{t}$, where
$T_{jk}=\cE_{jk}$ (for $j,k=1,2,3$) and $t_j=\cE_{j0}$. The
translation vector $\vec{t}$ describing the shift of the Bloch sphere
(including its center, i.e. the total mixture)
is related to the unitality of the channel. For unital maps $\vec{t}=\vec{0}$.

Diagonal elements of the state $\varrho$ are in this case associated
with the mean value $z=\T[\varrho S_3]$. The conservation of
the diagonal elements implies that the corresponding components of $\varrho$
are preserved. Combining the unitality with this property
we find the following form for  decoherence maps
\begin{equation}
\label{mm}
\cE=\left(
\ba{cccc}
1 & 0 & 0 & 0 \\
0 & a & b & 0 \\
0 & c & d & 0 \\
0 & 0 & 0 & 1 \\
\ea
\right)\; ,
\end{equation}
from where it follows
that the set of all possible qubit decoherence maps
is at most four-parametric.

Each unital map can be written as \cite{ruskai}
\be
\cE[\varrho]=R_{U_1} \Phi_\cE R_{U_2} [\varrho]=
U_1\Phi_\cE[U_2\varrho U_2^\dagger] U_1^\dagger\; ,
\ee
where $R_{U_1},R_{U_2}$ are orthogonal rotations
corresponding to unitary transformations $U_1,U_2$;
 $\Phi_\cE={\rm diag}\{1,\lambda_1,\lambda_2,\lambda_3\}$
and $\lambda_j$ are the singular values of the matrix $\cE$.
In fact, the above relation is the singular-value decomposition of the
matrix $\cE$. The conditions of the complete positivity restricts the
possible values of $\lambda_j$. In particular, the allowed
points $\vec{\lambda}=(\lambda_1,\lambda_2,\lambda_3)$
must lie inside a tetrahedron with vertices that have coordinates
$(1,1,1)$, $(1,-1,-1)$, $(-1,1,-1)$, and $(-1,-1,1)$, respectively.

Applying these facts to the decoherence
map under consideration ($\cE$ from Eq.(\ref{mm}))
we obtain that $\Phi_\cE={\rm diag}\{1,\lambda_1,\lambda_2,1\}$, i.e.
$\lambda_3=1$. Let us note that in this case we use unitaries that do not
change the decoherence basis, so we are still dealing with  all
decoherences that belong to a fixed basis $\cB$.
The condition of complete positivity restricts the values to the points
$\vec{\lambda}=(\lambda,\lambda,1)$ with $-1\le \lambda\le 1$,
i.e. to a line connecting the two vertices of the tetrahedron representing
the identity ($\lambda=1$) and the unitary rotation $S_3$
($\lambda=-1$). Consequently, the general decoherence channel
$\cE\in\cD_\cB$ reads
\begin{equation}
\nonumber
\cE=
\left(
\ba{cccc}
1 & 0 & 0 & 0 \\
0 & c_1 & s_1 & 0 \\
0 & -s_1 & c_1 & 0 \\
0 & 0 & 0 & 1 \\
\ea
\right)
\left(
\ba{cccc}
1 & 0 & 0 & 0 \\
0 & \lambda & 0 & 0 \\
0 & 0 & \lambda & 0 \\
0 & 0 & 0 & 1 \\
\ea
\right)
\left(
\ba{cccc}
1 & 0 & 0 & 0 \\
0 & c_2 & s_2 & 0 \\
0 & -s_2 & c_2 & 0 \\
0 & 0 & 0 & 1 \\
\ea
\right)\; ,
\end{equation}
where $s_j=\sin\varphi_j$ and $c_j=\cos\varphi_j$ represent rotations $R_{U_j}$
around the $z$-axis by an angle $\varphi_j$.
>From here it follows that a general decoherence map $\cE$ takes the form
\begin{equation}
\cE=\left(
\ba{cccc}
1 & 0 & 0 & 0 \\
0 & \lambda \cos(\varphi_1+\varphi_2) & \lambda\sin(\varphi_1+\varphi_2) & 0 \\
0 & -\lambda\sin(\varphi_1+\varphi_2)  & \lambda\cos(\varphi_1+\varphi_2)  & 0 \\
0 & 0 & 0 & 1 \\
\ea
\right)
\end{equation}
and, consequently, it is specified only by two real parameters
$a=\lambda\cos(\varphi_1+\varphi_2)$ and $b=\lambda\sin(\varphi_1+\varphi_2)$,
i.e.
\begin{equation}
\label{dec}
\cE=\left(
\ba{cccc}
1 & 0 & 0 & 0 \\
0 & a & b & 0 \\
0 & -b & a & 0 \\
0 & 0 & 0 & 1 \\
\ea
\right)\; .
\end{equation}
As a result we obtain that any map $\cE$ of the above form
with the numbers $a,b$ satisfying the condition $a^2+b^2\le 1$
is completely positive. Therefore we can conclude that the set of all
decoherence maps of a qubit
is characterized just by two parameters. Moreover, to obtain the decoherence
(to secure the suppression of off-diagonal terms) the inequality must be strict,
i.e. $a^2+b^2<1$. Otherwise the map $\cE$ desribes a unitary
rotation around the $z$ axis. Defining the rotation map
\be
\label{rotation}
R_\varphi=
\left(\ba{cc}
\cos\varphi & \sin\varphi\\
-\sin\varphi & \cos\varphi
\ea\right)
\ee
and using the relation $\varphi=\varphi_1+\varphi_2$, we can write
the most general decoherence channel ($\cE\in\cD_\cB$) in a very compact form
\be
\label{decoherence}
\cE=
\left(\ba{ccc}
1 & 0\ \ \ 0 & 0 \\
\ba{c} 0\\ 0 \ea & \lambda R_\varphi & \ba{c} 0\\ 0 \ea\\
0 & 0\ \ \ 0 & 1
\ea\right)\; .
\ee
This form is suitable for our purposes, because the powers of the map
$\cE$ read
\be
\label{discrete}
\cE^n=
\left(\ba{ccc}
1 & 0\ \ \ 0 & 0 \\
\ba{c} 0\\ 0 \ea & \lambda^n R_{n\varphi} & \ba{c} 0\\ 0 \ea\\
0 & 0\ \ \ 0 & 1
\ea\right)\; .
\ee

\section{Structural properties of decoherence channels}
In this section we will briefly review structural properties
of the set of all possible decoherence completely positive maps $\cE$.
Let us denote this set by $\cD$.
\begin{itemize}
\item{\bf Convex structure}\newline
The set of all decoherence maps $\cD$ is not convex, i.e. a convex combination
of two decoherence channels $\cE_\mu=\mu\cE_1+(1-\mu)\cE_2$
is not again a decoherence channel. This is true except the case when the
decoherence bases of $\cE_1, \cE_2$
coincide, i.e. the set $\cD_\cB$ is convex.
The extremal points of $\cD_\cB$ correspond to
unitary transformations. However, these are not elements
of $\cD_\cB$, because they do not fulfill the second
decoherence condition (2.2).
\begin{figure}
\begin{center}
\includegraphics[width=5cm]{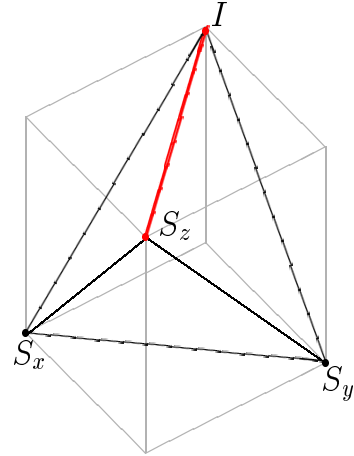}
\medskip
\caption{(Color online)
The cube corresponds to all positive unital trace-preserving maps.
The condition of complete positivity confines quantum channels into
the tetrahedron with (generalized) Pauli matrices as vertices.
In this picture the set of decoherence channels $\cD_\cB$
forms a line connecting the points $I$ and $S_z$.
}
\end{center}
\end{figure}
We have already mentioned that for qubits the set of all possible
$\Phi_\cE$ channels form a tetrahedron and up to unitary
trasnformations each channel belongs to this tetrahedron.
Those channels that correspond to decoherence maps form a line
connecting the points $(1,1,1)$ and $(-1,-1,1)$. From this picture
(see Fig.~1) the convexity of $\cD_\cB$ is transparent and also the extremal
points can be easily identified
as unitary channels. It follows that each decoherence map can be written
as a convex sum of only two unitary channels. In fact all maps
$\Phi_\cE$ for which one of the $\lambda$'s equals to unity and all the others
are the same define a decoherence with respect to some basis. This means that
all edges of the tetrahedron correspond to decoherence channels. It
illustrates that the set $\cD$ as a whole is not convex, but is composed
of a continuous number of ``convex'' subsets $\cD_\cB$ corresponding to each
orthonormal basis $\cB$.

\item{\bf Composition}\newline
A composition of two decoherence channels
$\cE=\cE_1\circ\cE_2$ is not, in general, a decoherence channel.
So the set $\cD$ is not closed under the operation of multiplication. The channel
$\cE$ belongs to $\cD$ only if the decoherence bases of
$\cE_1$ and $\cE_2$ coincide, i.e. again only the sets  $\cD_\cB$
are closed under the composition.

\item{\bf Classical capacity}\newline
The decoherence basis is preserved by the decoherence map. Therefore it
is possible to exploit these bases states to transmit
the maximally possible amount of information, i.e.
the capacity achieves its maximum $C=\log_2{d}$ with $d={\rm dim}\cH$.

\item{\bf Tensor product}\newline
The tensor product of two decoherence maps describes a decoherence.
However, $\cD_{12}\ne \cD_1\otimes\cD_2$, because the decoherence
basis of $\cE_1\otimes\cE_2$ is always separable. The open problem is
whether the whole set $\cD_{12}$ can be obtained
from the sets $\cD_1,\cD_2$ by global unitary rotations.
Properties of decoherence channels under tensor products is an
interesting topic, which is
related to our ability of controlling the
decoherence. For example, how the decoherence of a
sub-system affects characteristics of the whole system?
\end{itemize}

\section{Collision model}
In what follows we will study whether an arbitrary decoherence channel can be implemented
via a sequence of bi-partite collisions. Each of the collisions is described by a unitary transformation
$U$. Our task will be to derive all possible unitary transformations
that force the system to decohere. Our analysis  will
be performed only for qubits, but up to technical details all results
hold for qudits.

Let us consider that initially the system qubit is
decoupled from an environment (reservoir) that is modelled as a set of qubits, i.e.
$\Omega_{in}=\varrho\otimes \Xi_{res}$. Moreover, we will simplify the model
by assuming that initially the reservoir qubits are in a factorized state
$\Xi_{res}=\xi^{\otimes N}$ and each reservoir qubit interacts with the system qubit
just once. In addition we assume that reservoir qubits do not interact between themselves.
Under such conditions the evolution of the system
qubit is induced by the sequence of maps $\cE_1=\dots =\cE_N\equiv \cE$.
In particular, the state of the system after the $n$-th
interaction equals to
\be
\varrho^{(n)}=\cE_n\dots \cE_1[\varrho]=\cE^n[\varrho]\; ,
\ee
where $\cE[\varrho]=\T_{res}[U(\varrho\otimes\xi)U^\dagger]$. We will refer to this picture as to
to a collision model. The system qubit collides with reservoir qubits.

In order to obtain the decoherence channel, i.e.
$$
\varrho\to\varrho^{(n)}=\left(\ba{cc}
\varrho_{00} & \varrho_{12}^{(n)}\\
\varrho_{21}^{(n)} & \varrho_{11}
\ea\right)
$$
with $\varrho_{12}^{(n)}=[\varrho_{21}^{(n)}]^*\to 0$ for $n$ goes to infinity,
we have to ensure that
the map $\cE$ preserves  diagonal elements of each state
$\varrho$ in a given (decoherence) basis.

In order to preserve the diagonal elements
of pure states $|0\r\l 0|$ and $|1\r\l 1|$ (decoherence basis)
the bi-partite unitary transformation $U$
must necessarily satisfy the relations
\be
\nonumber |00\r &\to & |0\psi\r\; ; \\
\nonumber |01\r &\to & |0\psi^\perp\r\; ; \\
\nonumber |10\r &\to & |1\phi^\perp\r \; ;\\
 |11\r &\to & |1\phi\r \; .
\label{cond1}
\ee

In what follows we will prove our main result that the class of possible
bi-partite interactions that induce decoherence in collision models
coincides with the set of all {\it controlled-U transformations}
(the so-called U-processors as introduced in Ref.~\cite{hillery}),
where the system under consideration plays the role of the control and the reservoir particle is a target.
Certainly,
we have to identify those transformations for which the off-diagonal
elements of the system density operator do
vanish in the limit of infinitely many collisions with reservoir particles.

The unitary bi-partite transformation (the controlled-$U$ operation)
defined by the relations (\ref{cond1}) can be rewritten
into the following operator form
\be
U=|0\r\l 0|\otimes V_0+|1\r\l 1| \otimes V_1\; ,
\label{18}
\ee
where $V_0,V_1$ are unitary rotations of a reservoir qubit. In particular,
$V_0=|\psi\r\l 0|+|\psi^\perp\r\l 1|$ and
$V_1=|\phi^\perp\r\l 0|+|\phi\r\l 1|$.
Thus, the initial state $\Omega=\varrho\otimes\xi$ of a bi-partite system evolves according
to a transformation
\be
\Omega\to \Omega^\prime=U\Omega U^\dagger=
\sum_{j,k=0}^1\varrho_{jk}|j\r\l k|\otimes V_j\xi V_k^\dagger
\ee
and by performing the partial trace over the reservoir qubit we obtain the induced map
\be
\nonumber
\varrho\to\varrho^\prime&=&\cE[\varrho]=\T_p\Omega^\prime=
\sum_{j,k=0}^1\varrho_{jk} \T[V_j \xi V_k^\dagger] |j\r\l k|\\
\nonumber
&=& {\rm diag}[\varrho]+\varrho_{01} \l X\r_\xi |0\r\l\ 1|+\varrho_{10} \l X^\dagger\r_\xi |1\r\l 0|\; ,
\ee
where $X=V_1^\dagger V_0$ and $\l X\r_\xi=\T[X\xi]$ stands for the mean value
of the operator $X$ in the state $\xi$.

Applying the transformation $\cE$ in a sequence of $n$ collisions
the state of the system qubit is described by the density operator
\be
\nonumber
\varrho^{(n)}=\cE^n[\varrho]={\rm diag}[\varrho]+\varrho_{01} \l X\r_\xi^n |0\r\l 1|+
\varrho_{10} \l X^\dagger\r_\xi^n |1\r\l 0|\; ,
\ee
from where we can conclude, that providing $|\l X\r_\xi|< 1$
and $|\l X^\dagger\r_\xi|< 1$ the off-diagonal terms vanish. However,
because $XX^\dagger=X^\dagger X=\openone$, i.e. $X$ is unitary, its
eigenvalues are just complex square roots of the unity. Therefore, for
the eigenvectors of $X$
the off-diagonal terms do not tend to zero.

The fact that for  convex combinations of the eigenvectors
the off-diagonal elements still vanish might sound counterintuitive.
But it can be seen from the following consideration:
Let us denote by $e^{i\varphi}$ and $e^{i\eta}$ the eigenvalues of $X$
associated with the eigenvectors $|f_1\r$ and $|f_2\r$, respectively.
Then the mean value $\l X\r_\xi$ for the convex combination
$\xi=a|f_1\r\l f_1|+(1-a)|f_2\r\l f_2|$ equals to
$\l X\r_\xi=e^{i\varphi} a+e^{i\eta} (1-a)$. The condition
$|\l X\r_\xi|<1$ can be rewritten as the inequality
$2a(1-a)[1-\cos(\varphi-\eta)]<0$, which is satisfied only if
$\cos(\varphi-\eta)\ne 1$, or $a\ne 1$ and $a\ne 0$. The latter property means that
$\xi$ is the eigenstate. The first property requires $\varphi=\eta$,
i.e. the operator $X$ is proportional to the identity,
$X=e^{i\varphi}\openone$. However, under this assumption
$V_1=e^{i\varphi} V_0$, i.e. we have no interaction and
$U=(e^{i\varphi}|0\r\l 0|+|1\r\l 1|)\otimes V_1$.
Hence, we can conclude that whenever the reservoir state is not an
eigenstate of $X$ and the interaction is not trivial,
the described collision model with controlled-$U$ interaction
forces the system to decohere.

It is straightforward to show that
 unitary interactions $U=|0\r\l 0|\otimes V_0+|1\r\l 1|\otimes V_1$
induce maps of the left-right form (see Eq.(\ref{dec}))
with the parameters
\be
\nonumber
a&=& \frac{1}{2}(\l X\r_\xi+\l X^\dagger\r_\xi)\; ;\\
b&=& \frac{i}{2}(\l X\r_\xi-\l X^\dagger\r_\xi)\; ,
\ee
or, equivalently, $\l X\r_\xi = \lambda e^{i\varphi}$. So given a decoherence
map $\cE$ one can, in principle, find an interaction $U$
and an initial state of the reservoir qubits $\xi$,
such that the desired decoherence process is implemented via a sequence of collisions.

\section{Master equation}
In this section we will derive a master equation that describe the decoherence
process induced by collisions of the system qubit
with reservoir particles. Although the studied decoherence model
is intrinsically discrete, we will show that we can perform a continuous-time
approximation that enable us to write down the master equation (see, e.g. \cite{ziman_torun}).

As shown in the previous section the collision model is
described by a set of maps $\cE_n=\cE^n$ that form a discrete
semigroup, i.e. $\cE_n \cE_m = \cE_{n+m}$ for all integer
$m,n$ and $\cE_0=\cI$. The question is whether we can introduce
a continuous one-parametric set of transformations $\cE_t$
such that $\cE_{t_n}=\cE_n$ for $t_n = n\tau$ ($\tau$ is a time scale
roughly corresponding to the time interval between two interactions).
It turns out that a simple relation $n\to t/\tau$ can be used to accomplish the task.
The obtained continuous set of transformations $\cE_t$
will be used to derive the generator $\cG$ of the dynamics
by using a simple formula $\cG_t=\dot\cE_t \cE_t^{-1}$.

With the help of results from Sec.~III (namely, Eq.(\ref{discrete}))
we can directly write
\be
\cE_t=\left(
\ba{ccc}
1 & 0 \ \ \ 0 & 0 \\
\ba{c} 0 \\ 0 \ea &
\lambda^t R_{t\varphi}
& \ba{c} 0 \\ 0 \ea \\
0 & 0 \ \ \ 0 & 1
\ea
\right)\, ,
\ee
where for simplicity we set the time scale $\tau=1$.
It is easy to see that the one-parametric set
of transformations $\cE_t$ possesses the semigroup property, i.e.
$\cE_t \cE_s = \cE_{t+s}$. for all real $t,s$. It means that the generator
and the associated master equation will be of the Lindblad form
\cite{spohn}, i.e. the process under consideration is Markovian.

The corresponding generator reads
\begin{equation}
\cG=\left(
\ba{cccc}
0 & 0 & 0 & 0 \\
0 & \ln\lambda & -\varphi & 0 \\
0 & \varphi & \ln\lambda & 0 \\
0 & 0 & 0 & 0
\ea
\right)\; ,
\end{equation}
where we used the identity $\dot R_{t\varphi}=\varphi R_{t\varphi+\pi/2}$
and
$\frac{d}{dt}[\lambda^t R_{t\varphi}]\lambda^{-t}R_{-t\varphi}
=\ln\lambda R_0+\varphi R_{\pi/2}$. This step
can be performed only if $\lambda$ is non-negative (i.e., when the logarithm is defined),
which, in general, is
not the case. The parameter $\lambda$ belongs to the open interval
$(-1,1)$. Consequently, it seems that
the generator cannot be derived in all cases.
However, using the equality $-\lambda R_{\varphi}=\lambda R_{(\varphi+\pi)}$
for $\lambda$ nonnegative, one can write
$|\lambda|^t R_{t(\varphi+\pi)}$ instead of $\lambda_t R_{t\varphi}$
in the expression for $\cE_t$ with $\lambda<0$.
Then the generator is slightly different
and contains the term $\varphi+\pi$ instead of $\varphi$, and
$\ln|\lambda|$ instead of $\ln\lambda$. Thought this is  not a problem,
because in terms of parameters of the collision model
$\l X\r_\xi=\lambda e^{i\varphi}$, i.e. the parameter
$\lambda=|\l X\r_\xi|$ is always
positive. Therefore we can consider the generator $\cG$ as the most general one.

The general master equation in Lindblad form reads
\be
\nonumber
\dot\varrho_t =\cG[\varrho_t]=-i[H,\varrho_t]+
\frac{1}{2}\sum_{a,b} c_{ab}([S_a,\varrho_t S_b]+[S_a\varrho_t, S_b])\; .
\ee
If the numbers $c_{ab}$ are time-independent and form a positive matrix,
then the generated evolution is Markovian and satisfies the semigroup
property. To find the values of the coefficients $c_{ab}$ we will
use the following relations (see Ref.\cite{ziman_torun})
\be
\nonumber
h_1=\frac{[\cG]_{32}-[\cG]_{23}}{4}\; ; &
h_2=\frac{[\cG]_{13}-[\cG]_{31}}{4}\; ; &
h_3=\frac{[\cG]_{21}-[\cG]_{12}}{4}\; ; \\
e_{23}=\frac{[\cG]_{10}}{4} ; &
e_{31}=\frac{[\cG]_{20}}{4}; &
e_{12}=\frac{[\cG]_{30}}{4};
\ee
and
\be
\nonumber
d_{11}=\frac{[\cG]_{11}-[\cG]_{22}-[\cG]_{33}}{4}\; ; \; &
d_{12}=\frac{[\cG]_{12}+[\cG]_{21}}{4}\; ; \\
\nonumber
d_{22}=\frac{[\cG]_{22}-[\cG]_{11}-[\cG]_{33}}{4}\; ; \; &
d_{23}=\frac{[\cG]_{23}+[\cG]_{32}}{4}\; ; \\
\nonumber
d_{33}=\frac{[\cG]_{33}-[\cG]_{11}-[\cG]_{22}}{4}\; ; \; &
d_{13}=\frac{[\cG]_{13}+[\cG]_{31}}{4}\; , \\
\ee
where $[\cG]_{kl}$ correspond to matrix elements of the generator $\cG$,
$c_{ab}=d_{ab}-ie_{ab}$ and $H=\sum_a h_a S_a$. Note that $d_{ab}$
form a symmetric matrix and $e_{ab}$ is an antisymmetric matrix.

Using these expressions one finds that the non-vanishing
parameters are
\be
h_3=\frac{1}{2}\varphi\; ; \ \ \ \ \ \
d_{33}= - \frac{1}{2}\ln\lambda
\ee
and the corresponding master equation reads
\be
\label{dme}
\dot\varrho_t = -i\frac{\varphi}{2}[S_3,\varrho_t]-\frac{\ln\lambda}{2}
(S_3\varrho_t S_3 - \varrho_t)\; .
\ee
A typical evolution driven by this equation is depicted
in Fig.~2.
\begin{figure}
\includegraphics[width=8cm]{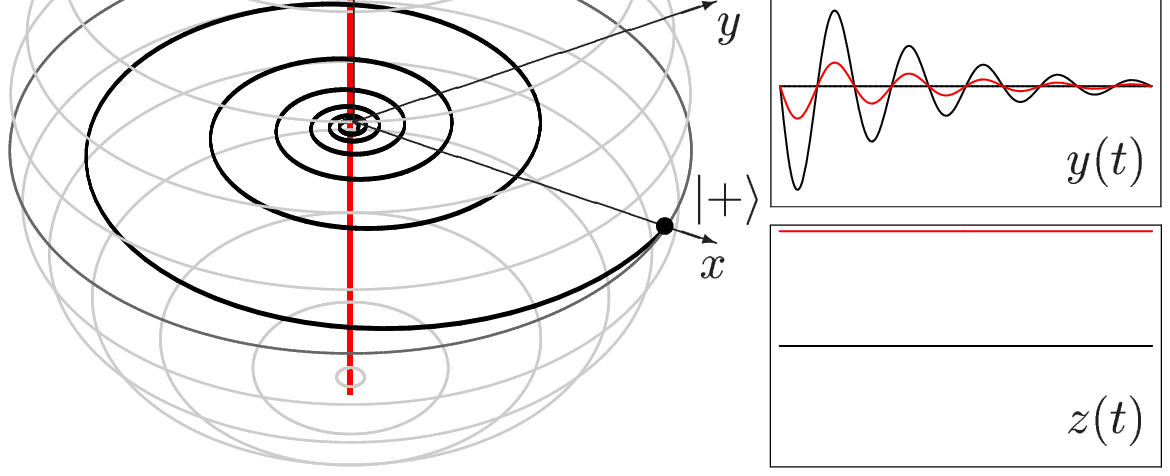}
\medskip
\caption{
(Color online) The decoherence of a qubit governed by Eq.(\ref{dme}).
The Bloch sphere that represents the initial state space of a qubit is mapped into the line connecting
the decoherence basis states. On the right the evolution of
Bloch-vector components for two different initial states is depicted.
}
\end{figure}

Let us now address the following question: Is there any other master
equation describing a decoherence of a qubit? The preservation of the
$S_z$ component (determined by the decoherence basis) together with
the unitality of the transformation implies that
\begin{equation}
\cG=\left(
\ba{cccc}
0 & 0 & 0 & 0 \\
0 & a & b & 0 \\
0 & c & d & 0 \\
0 & 0 & 0 & 0
\ea
\right)\; .
\end{equation}
The corresponding matrix $C=\frac{1}{2}[c_{ab}]$ then reads
\begin{equation}
C=\frac{1}{4}\left(
\ba{ccc}
a-d & c+b & 0 \\
c+b & d-a & 0 \\
0 & 0 & -a-d \\
\ea
\right)\; .
\end{equation}
This matrix is positive only when
$a=d$ and $b=-c$. Moreover, $a$ must be negative.
These restrictions leave
only a single element that does not vanish, namely, $c_{33}=-a/2$. Consequently,
the Hamiltonian part takes non-vanishing value for $h_3=b/2$. Therefore
the family of {\em all} master equations describing the decoherence is only
two-parametric
\be
\label{gme}
\dot\varrho_t = -i\frac{b}{2}[S_3,\varrho_t]-\frac{a}{2}
(S_3\varrho_t S_3 - \varrho_t)\; .
\ee
This general master equation is of the same form
as the one derived for the collision model (5.6).
The parameters $\lambda,\varphi$ are related
to the parameters of the underlying unitary interaction
via the formula $\l X\r_\xi=\lambda e^{i\varphi}$. Let us note the constraint
$\lambda=|\l X\r_\xi|\in [0,1]$, since $X$ is unitary. Therefore
$\ln\lambda\le 0$ as it is required by the condition on possible
values of $a$.

\section{Entanglement in decoherence via collisions}

We start with definitions of entanglement quantities that
we will evaluate. Let us denote the joint state of the system of $N+1$
qubits (the system qubit and $N$ reservoir qubits) by $\Omega$.
The bipartite entanglement shared between a pair of qubits $j$ and $k$ can be quantified in
terms of the concurrence \cite{wootters}
\be
C_{jk} = \max\{0,\lambda_1-\lambda_2-\lambda_3-\lambda_4\}\; ,
\ee
where $\lambda_j$ are decreasingly ordered square roots of the eigenvalues
of the matrix $R_{jk}=\varrho_{jk}
\sigma_y\otimes\sigma_y \varrho^*_{jk}\sigma_y\otimes\sigma_y$ and
$\varrho_{jk}=\T_{\overline{jk}}\Omega$ is the state of two qubits
under consideration.

The case of multi-partite entanglement is a more complex phenomenon
and there is no unique way of its quantification. Fortunately, for
{\em pure} multi-qubit systems there is an accepted method of
characterization (identificantion) of intrinsic multi-partite entanglement.
Specifically, let us consider how strongly the $j$-th qubit is correlated with the
rest of qubits in the multi-partite system. This degree of entanglement can be quantified
via the so-called tangle (see Ref.~ \cite{ckw})
\be
\tau_j=4\det\varrho_j=2(1-\T\varrho_j^2)\; ,
\ee
where $\varrho_j=\T_{\overline{j}}\Omega$ is the state
of the $j$-th qubit. Then we evaluate bi-partite concurrences between the given $j$-th qubit
and any other qubit in the system, i.e. we evaluate $N$ quantities $C_{jk}$.

Wootters and his coworkers have found (see Ref.~\cite{ckw}) that
for pure three-qubit states the inequalities
\be
\sum_{j\ne k} [C_{kj}]^2 \le \tau_k\; \qquad \forall k=1,2,3 \; ,
\ee
hold. In addition they have conjectured
that such inequalities also hold for any number of qubits. This conjecture
(to so-called Coffman-Kundu-Wootters (CKW) inequality) has been recently
proved by Osborne \cite{osborne2005}
These inequalities quantify the property which is known as
the {\em monogamy of entanglement} (the entanglement
cannot be shared freely in multipartite systems).

As a consequence of the CKW inequality one can define a measure
of intrinsic multipartite entanglement $\Delta_j$ as
\be
\Delta_j = \tau_j - \sum_{k\ne j} \tau_{jk}\; ,
\ee
where we have used the notation $\tau_{jk}=[C_{jk}]^2$.
It is important to note that in the multi-partite case
(in particular for more than three qubits) the differences
$\Delta_j:=\tau_k - \sum_{j\ne k} \tau_{jk}$ take different values
for different $j$. Therefore,
a weighted sum $\Delta=\frac{1}{N} \sum_j \Delta_j$ is an appropriate
measure of an intrinsic multipartite entanglement. Based on
this quantity we can argue
that there are multi-partite entangled states for which
the entanglement has purely bi-partite origin, as for example the family
of $W$ states \cite{ziman_ckw} that saturate the
CKW inequalities, i.e. $\Delta=0$.

Let us assume that the system qubit is initially prepared in
the state $|\chi\r=a|0\r+b|1\r$ and each qubit of the reservoir is in a
pure state $|\psi\r$, i.e. the joint initial state is
$|\Omega_0\r=|\chi\r\otimes |\psi\r^{\otimes N}$. After $n$ collisions
governed by bi-partite controlled unitary operations (\ref{18})
the whole system evolves into the state
\be
\label{omega_n}
|\Omega_n\r=
\left[a|0\r\otimes|V_0\psi\r^{\otimes n}+
b|1\r\otimes |V_1\psi\r^{\otimes n}
\right]\otimes|\psi\r^{(N-n)}\; .
\ee
In order to be able to evaluate the entanglement quantities
we have to specify all two-qubits and single-qubit density operatos. In particular,
for $k\le n$, $j\le k$ the bi-partite states are given by expressions
\be
\nonumber
\varrho_{0k}(n)&=&|a|^2 |0\psi_0\r\l 0\psi_0|+|b|^2|1\psi_1\r\l 1\psi_1|\\
& & +ab^*|\l \psi_0|\psi_1\r|^{(n-1)}|0\psi_0\r\l 1\psi_1|+c.c. \; ;\\
\varrho_{jk} (n)&=& |a|^2 |\psi_0\psi_0\r\l\psi_0\psi_0|+|b|^2
|\psi_1\psi_1\r\l\psi_1\psi_1|\; '
\ee
where we used the notation $|\psi_0\r=V_0|\psi\r$ and
$|\psi_1\r=V_1|\psi\r$.
The single qubit states are as follows:
\be
\nonumber
\varrho_0 (n)=|a|^2 |0\r\l 0|+|b|^2|1\r\l 1|
+ab^*|\l \psi_0|\psi_1\r|^{n}|0\r\l 1|+c.c.
\ee
describes the system qubit after $n$-th collision, and
\be
\label{xi_n}
\varrho_k(n)= |a|^2 |\psi_0\r\l\psi_0|+|b|^2
|\psi_1\r\l\psi_1|
\ee
describes the $k$-th qubit of the reservoir after the collision with the system qubit.
Evaluation of the tangles is straightforward and results in expressions
\be
\tau_0(n)&=& 4 |a|^2 |b|^2 (1-|\l\psi_0|\psi_1\r|^{2n})\; ; \\
\tau_k(n)&=& 4 |a|^2 |b|^2 |\l\psi_0|\psi_1^{\perp}\r|^2 \; ;\\
\tau_{0k}(n) &=& 4 |a|^2 |b|^2 |\l\psi_0|\psi_1\r|^{2(n-1)} |\l\psi_0|\psi_1^\perp\r|^2 \; ;\\
\tau_{jk}(n)&=& 0\; .
\ee
One can directly verify the validity of the CKW inequalities
\be\nonumber
\sum_{k=0,k\ne j}^N \tau_{jk}(n) &=& \tau_{j0}(n)=
4 |ab|^2 |\l\psi_0|\psi_1\r|^{2(n-1)}|\l\psi_0|\psi_1^\perp\r|^2\\
&\le & 4 |ab|^2 |\l\psi_0|\psi_1^\perp\r|^2 = \tau_j(n)
\; ;\\
\nonumber
\sum_{k=1}^N \tau_{0k}(n)&=&n\times 4|ab|^2 |\l\psi_0|\psi_1\r|^{2(n-1)}|\l\psi_0|\psi_1^\perp\r|^2 \\
&\le& 4 |ab|^2 (1-|\l\psi_0|\psi_1\r|^{2n}) = \tau_0(n)
\; ,\ee
where we have used the relations $|\l\psi_0|\psi_1\r|\le 1$
and $|\l\psi_0|\psi_1^\perp\r|^2=1-|\l\psi_0|\psi_1\r|^2$.

\begin{figure}
\includegraphics[width=8cm]{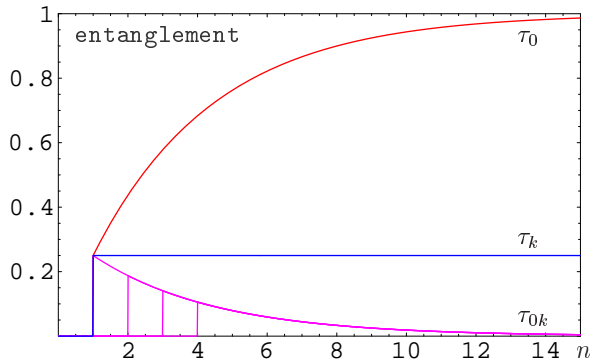}
\medskip
\caption{(Color online)
The behavior of entanglement as a function of number $n$ of
collisions between the system qubit and reservoir qubits.
The degree of entanglement between the system qubit and $n$ reservoir qubits
after $n$ collisions is given by
$\tau_0$  - it
increases with the number of collisions (time) to a steady-state value.
On the contrary, all reservoir qubits after their interaction with the system qubit
are entangled with the constant degree of entanglement (see the tangle $\tau_k$).
The bi-partite entanglement
$\tau_{0k}$ (the square of the concurrence $C_{0k}$)
is zero until the $k$-th reservoir qubit collides with the system qubit.
After the collision the entanglement takes a non-zero value, though it decreases due to
subsequent collisions of the system qubit with other reservoir qubits.
It is interesting to note that all $\tau_{0k}(n)$ for $n\ge k$ are described
by the same function. We assune the following initial state of the system qubit
 $|\psi\r=\frac{1}{\sqrt{2}}(|0\r+|1\r)$ and $|\l\psi_0|\psi_1\r|^2=0.75$.
}
\end{figure}

In the limit of large number of interactions ($n\to\infty$) all two-qubit
correlations  vanish (i.e. finally there is no bi-partite entanglement
between qubits in reservoir), but the entanglement between the
system qubit and the whole reservoir converges to a finite value
\be
\tau_0 &\to& 4 |ab|^2 \; ;\\
\tau_{0k} & \to & 0\; .
\ee
It means that after the process of decoherence the system qubit
is not entangled with the reservoir via bipartite entanglements, but
is entangled to the reservoir via multi-partite correlations. The final
state belongs to the family of Greenbeger-Horn-Zeilinger states that exhibit
purely multi-partite correlations.

>From the above one can see how the entanglement is
related to the decoherence. Given the relation
$|\l\psi_0|\psi_1\r|=|\l X\r_\psi|=\lambda$ we conclude that the decoherence rate
 restricts the maximum amount of created entanglement and
simultaneously it determines the decrease of entanglement with the number of
collisions.

\section{Summary and conclusions}

In this paper we have studied qubit decoherence channels as defined by Eq.~(\ref{1.1}).
We have presented their complete classification. In addition, we have shown that all
decoherence channels can be modelled as collisions of a quantum system with its environment.
The bi-partite collisions between the system and reservoir particles are modelled as
the controlled-$U$ operations such that the system particle is a control while a reservoir
particle is a target. Using the collision model we have derived the most general decoherence master equation
in the Lindblad form that describes decoherence. The specific basis in which the decoherence
takes place as well as the decoherence rates are specified by properties of the controlled-$U$
operation and the initial state of reservoir particles. We have shown that in the collision model
the decoherence is accompanied (or, from a different point of view, one can say that the decoherence is due to)
quantum entanglement that is created between the system particle and the reservoir particles. We have derived
the explicit expressions for entanglement measures
(the concurrence between an arbitrary pair of particles involved in the
dynamics and a tangle that characterize a degree of entanglement between the given particle and the rest of
the system). Using these measures and the Coffman-Kundu-Wootters inequalities we have shown that in the case
of decohering qubit collisions between this qubit and the reservoir lead to intrinsic multi-qubit entanglement
of all qubits involved in the process.

We conclude our paper with some remarks.

{\it i)}\newline
Even though through the paper we have been paying attention mostly to decoherence
of qubits many of the results
hold in general. In particular, within the framework of
a collision model with the controlled-$U$ bi-partite collisions
(the system particle plays the role of the control while particles from the reservoir are targets)
a decoherence of qudits can be described as well.

{\it ii)}\newline
The collision model used in this paper is a discrete one. We have assumed that a collision between two particles
is localized in time, so that at a given time instant the controlled-$U$ operation (a bi-partite gate) is applied.
The sequence of interactions is then labelled by an integer number $n$ and the total dynamics is represented
by a discrete semigroup.

As shown in the paper it is straightforward to introduce a continuous
time parameter so that the continuous evolution version of the sequence of collision is
described by a Markovian process represented by a continuous semigroup.
We have derived the corresponding master equation that describes the process of decoherence.
More importantly, we have shown that for qubits this
master equation describing the decoherence is unique and takes the form
 (\ref{gme}) that can be written as
\be
\label{deceq}
\dot{\varrho}_t = -i[H,\varrho_t]-\frac{1}{2\gamma}[H,[H,\varrho_t]] \, ,
\ee
where we use the notation $H= \frac{b}{2}S_3=\frac{\varphi}{2} S_3$
and $\gamma=-b^2/2a=\varphi^2/2\ln\lambda$. We note that the double
commutator term is well known and usually appears in decoherence models
even for higher-dimensional systems. For example,
Milburn in his work on intrinsic decoherence (see Ref.~\cite{milburn})
has derived a generalization of the usual
Schr\"odinger equation exactly in the form (\ref{deceq}).

{\it iii)}\newline
We have shown that the decoherence in the collision model is accompanied (caused)
by a creation of entanglement between the system and the reservoir.
Unlike in the process of homogenization described in \cite{ziman,ziman_ckw,ziman_phd},
in which the created entanglement saturates the CKW inequalities, in
the case of decoherence the entanglement results in the Greenberger-Horn-Zeilinger type of correlations
\cite{vidal}.
This means that decoherence process (as described by our collision model)
does not create an entanglement between the environment particles.
Specifically, if we trace over the system qubit (which decoheres)
in the $n$th step of the evolution (see Eq.(\ref{omega_n})),
we find that the environment is in a separable state
\be
\nonumber
\omega_{env} (n) &=& \T|\Omega_n\r\l \Omega_n| \\
\nonumber &=&
\left[|a|^2 (|\psi_0\r\l \psi_0|)^{\otimes n}+
|b|^2 (|\psi_1\r\l\psi_1|)^{\otimes n}\right]
\\ \nonumber & & \ \ \ \ \ \ \ \otimes\;
|\psi\r\l\psi|^{\otimes(N-n)}\; ,
\ee
where all the parameters are specified in Sec.~VI.
The decoherence rate $\lambda$ and the rotation parameter $\varphi$
can be adjusted by a suitable choice of the interaction $U$ and the state
of the reservoir $\xi$. The collision model reflects microscopic
origins of both these parameters that enter the decoherence master
equations. The eigenvalues of the Hamiltonian $H$
are given by the value of $\varphi$ and the parameter $\gamma$
is specified by both these parameters. The eigenvectors of $H$ form the decoherence
basis.

{\it iv)}\newline
We have shown explicitly that an arbitrary decoherence channel for a qubit can be represented
via the collision model with a particularly chosen controlled-$U$ interaction.
However, this result holds for arbitrary dimension (i.e. for qudits) as well.
Let remind us that an arbitary quantum
map $\cE$ can be represented as unitary operation
on some larger system (this is a content of the Stinespring-Kraus
dilation theorem \cite{nielsen}).
We have shown that for decoherence channels the collision (represented by a unitary transformation)
must be of the form of the controlled-$U$ operation. An open question is whether
each decoherence master equation (even for ${\rm dim}\cH=\infty$)
can be derived from the collision model. Knowing a decoherence master equation
(i.e., knowing a generator $\cG$) it is easy to ``fix'' a time step
$t=\tau$ and define $\cE_{\tau}=\cE$. This map is
for sure a decoherence channel and can be realized by a collision $U$. By applying this
``elementary'' map many times (a sequence of collisions)
we obtain a discrete semigroup of the powers of
$\cE$. The inverse task is trickier, that is, how do we
interpolate between these discrete sequence of transformations
(parameterized by number of collisions)
 to obtain a continuously parameterized channel. From a construction of the problem we
 know that the solution exists (we have started our analysis
  from the master equation).
The  question is whether this interpolation for qudit channels can
be performed as easily as for qubits, i.e. by replacing the discrete
powers of $n$ with continuous parameter $t$.
Nevertheless, given the fact that we have started
with a continuous set of channels $\cE_t$ and by replacing $t\to \tau$
we obtained $\cE_1=\cE$. Consequently, it is possible to replace
$n\to t/\tau$ to obtain the original continuous semigroup
of decoherence channels $\cE_t$. As a result we have found that a collision
model can be used not only to describe any decoherence master equation,
but can also be used to describe any quantum evolution governed by the Lindblad equation. On the other hand, it has to be stressed that
collision models describe evolutions that might not be
``interpolated'' by continuous semigroup of quantum channels \footnote{Mathematically,
this is related to the property of {\it infinite divisibility}
of the matrix $\cE$, i.e. to the possibility to calculate
all real powers.}.


\acknowledgements This work was supported in part by the European
Union  projects QGATES, QUPRODIS and CONQUEST,  by the Slovak
Academy of Sciences via the project CE-PI, by the project
APVT-99-012304 and by the Alexander von Humboldt Foundation.



\end{multicols}


\begin{thebibliography}{00}

\bibitem{Peres}
A.Perez: {\it Quantum Theory: Concepts and Methods},
(Kluwer, Dordrecht, 1993)

\bibitem{zeh}
E. Joos and  H.D.Zeh,
{\it The emergence of classical properties through interactions with the enviroment},
Z. Phys. B {\bf 59} 223 (1985)

\bibitem{zurek}
W.H. Zurek,
{\it Decoherence and the transition from quantum to classical},
Physics Today {\bf 44}, Num. 10, 36 (1991);
see also the revised version {\tt quant-ph/0306072}

\bibitem{zurek2003}
W.H. Zurek,
{\it Decoherence, einselection, and the quantum origins of the classical}
Rev. Mod. Phys. {\bf 75}, 715 (2003);
see also {\tt quant-ph/0105127}.

\bibitem{giulini}
D. Giulini, E. Joos, C. Kiefer, J. Kupsch, I.-O. Stamatescu, and H.D.Zeh,
{\it Decoherence and the Appearance of a Classical World in Quantum Theory},
(Springer, Berlin, 1996)

\bibitem{schlosshauser}
M. Schlosshauser,
{\it Decoherence, the measurement problem, and interpretations of quantum mechanics},
Rev. Mod. Phys. {\bf 76}, 1267 (2004);
see also {\tt quant-ph/0312059}

\bibitem{nielsen}
M.A. Nielsen and I.L. Chuang,
{\it Quantum Computation and Quantum Information},
(Cambridge University Press, Cambridge, 2000)

\bibitem{milburn}
G.Milburn,
{\it Intrinsic decoherence in quantum mechanics},
Phys. Rev. A {\bf 44}, 5401-5406 (1991)

\bibitem{haroche}
S. Haroche,
{\it Entanglement, mesoscopic superpositions and decoherence studies with atoms and photons in cavity},
Physica Scripta {\bf T76}, 159 (1998)

\bibitem{arndt}
K. Hornberger, S. Uttenthaler, B. Brezger, L.  Hackerm\"uller, M. Arndt, and A. Zeilinger,
{\it Collisional decoherence in matter wave interferometry},
Phys. Rev. Lett. {\bf 90}, 160401 (2003)

\bibitem{dennis}
D. Kretschmann and R.F. Werner,
{\it Quantum channels with memory},
{\tt quant-ph/0502106}

\bibitem{ruskai}
M.B. Ruskai, S. Szarek, and E. Werner,
{\it A characterizarion of completely positive tracepreserving maps on ${\cal M}_2$},
Lin. Alg. Appl. {\bf 347}, 159 (2002)

\bibitem{hillery}
M. Hillery, M. Ziman, and V. Bu\v zek,
{\it Implementation of quantum maps by programmable quantum processors},
Phys. Rev. A {\bf 66}, 042302 (2002)


\bibitem{ziman_torun}
M. Ziman, P. \v Stelmachovi\v c, and V. Bu\v zek,
{\it Description of quantum dynamics of open systems based on collision-like models},
to appear in Open Systems and Information Dynamics (2005)


\bibitem{spohn}
H. Spohn,
{\it Kinetic equations from Hamiltonian dynamics: Markovian limit},
Rev. Mod. Phys. {\bf 53}, 569 (1980)


\bibitem{wootters}
W.K. Wootters,
{\it Entanglement of formation of an arbitrary state of two qubits},
Phys. Rev. Lett. {\bf 80}, 2245 (1998)

\bibitem{ckw}
V. Coffmann, J. Kundu, and W.K. Wootters,
{\it Distributed entanglement},
Phys. Rev. A {\bf 61}, 052306 (2000)


\bibitem{osborne2005}
T.J. Osborne,
{\it General monogamy inequality for bipartite qubit entanglement},
{\tt quant-ph/0502176}

\bibitem{ziman}
M. Ziman, P. \v Stelmachovi\v c, V. Bu\v zek, M. Hillery, V. Scarani, and N.Gisin,
{\it Dilluting quantum information: An analysis of information transfer in system-reservoir interactions},
Phys. Rev A {\bf 65}, 042105 (2002),
see also {\tt quant-ph/0110164}

\bibitem{ziman_ckw}
M. Ziman, P. \v Stelmachovi\v c, and V. Bu\v zek,
{\it Quantum homogenization: Saturation of CKW inequalities},
J. Optics B: Quantum Semiclass {\bf 5}, 439 (2003)

\bibitem{ziman_phd}
M. Ziman,
{\it Entanglement as a structure: Application to quantum information processing}
(PhD thesis, Bratislava 2003)

\bibitem{vidal}
W. D\"ur, G. Vidal, and I. Cirac,
{\it Three qubits can be entangled in two inequivalent ways},
Phys. Rev. A {\bf 62}, 062314 (2000)


\end{thebibliography}
\end{document}